\documentclass[prl,twocolumn,showpacs,preprintnumbers,amsmath,amssymb]{revtex4}
\usepackage{colordvi}
\usepackage{amssymb}
\usepackage{graphics}
\usepackage{pstricks}
\usepackage{fancybox}
\usepackage{epsfig}

\newcommand{\gtap}{\;{\raise.3ex\hbox{$>$\kern-.75em\lower1ex\hbox{$\sim$}}}\;}
\newcommand{\ltap}{\;{\raise.3ex\hbox{$<$\kern-.75em\lower1ex\hbox{$\sim$}}}\;}

\begin{document}

\preprint{RM3-TH/0409, LAPTH-1043-04}

\title{Observation potential for $\eta_b$ at the Tevatron} 

\author{F. Maltoni} 
\email{maltoni@fis.uniroma3.it}
\affiliation{
Centro Studi e Ricerche ``Enrico Fermi'', 
via Panisperna, 89/A - 00184 Rome, Italy}
\author{A.D. Polosa}
\email{Antonio.Polosa@cern.ch}
\affiliation{
LAPTH, 9, Chemin de Bellevue, 
BP 110, 74941 Annecy-le-Vieux Cedex, France}

\date{\today}

\begin{abstract}
We calculate the cross section for $\eta_b$ production at the Tevatron
at next-to-leading order in the strong coupling and find 
that more than two millions of $\eta_b$'s are expected per inverse 
picobarn of integrated luminosity. We discuss the decay 
modes into charmed states and suggest that the decays into $D^* D^{(*)}$ 
mesons might be the most promising channels to observe the $\eta_b$ in Run~II.

\end{abstract}


\maketitle


Considering the rich phenomenology of the $\Upsilon$ states,
it is quite surprising that spin singlet $b \bar b$ states,
including the $^1S_0$ ground state, have not been observed yet.

Fine and hyperfine splittings of the quarkonia spectra were
calculated using phenomenological models for the
heavy-quark potential~\cite{Eichten:1980ms,Eichten:1994gt}.  Recent
progress both in lattice and in perturbative QCD has allowed the
achievement of comparable
precisions~\cite{Liao:2001yh,Recksiegel:2003fm}.  Various approaches
have been successfully adopted to describe the charmonium system and
are believed to be even more reliable for the heavier $b\bar b$ states,
where relativistic effects are less important. Recent determinations
lead to a mass splitting between the $\Upsilon(1S)$ and the
$\eta_b(1S)$ in the 40--60 MeV
range~\cite{Liao:2001yh,Recksiegel:2003fm,Ebert:2002pp,Lengyel:2000dk,
Kniehl:2003ap}.

Searches of the $\eta_b$ have been pursued in various experiments.  In
$e^- e^+$ collisions, cross sections for producing spin singlet states
are generally small and the signal is rate-limited. This is
compensated by a clean environment, which allows for searches in the
inclusive decay modes.  Following an original suggestion by Godfrey
and Rosner~\cite{Godfrey:2001eb}, the CLEO collaboration has looked
for the $\eta_b$ in the hindered M1 decay of the $\Upsilon(3S)$ and
found no signal~\cite{Mahmood:2002jd}. At LEP II, the ALEPH
collaboration analysed the $\gamma \gamma$ interactions data, basing
their analysis on the QCD prediction of $\Gamma_{\gamma\gamma}$
partial width~\cite{Czarnecki:2001zc}: no evidence was found in the
four- and six-charged-particle decay modes~\cite{Heister:2002if}.

The situation is exactly reversed in hadron collisions, where the
production rates can be large, with millions of events produced at the
Tevatron per inverse picobarn of integrated luminosity, yet the
intense hadronic activity makes any inclusive analysis unfeasible. In
this case it is necessary to identify decay modes that have
triggerable signatures and allow for full invariant mass
reconstruction of the decaying state.  Such exclusive modes are
believed to have very small branching ratios.  Braaten et
al.~\cite{Braaten:2000cm} have suggested that the $\eta_b$ at the
Tevatron could be observed via its decay into $J/\psi\, J/\psi$,
with the subsequent leptonic decay of the $J/\psi$'s; experimental 
efforts have started in this direction~\cite{Tseng:2003md}. This signature
exploits the upgraded abilities of the CDF detector of triggering on
soft muons.

The purpose of this note is twofold. First, we provide the
prediction for the inclusive cross section of $\eta_b$ production at
the Tevatron, based on the (NR)QCD calculation at the next-to-leading
accuracy in the strong coupling~\cite{Petrelli:1998ge}.  We find that
the cross section is about four times larger than the previous
available estimate~\cite{Braaten:2000cm}. 
Second, stemming from an estimate of the corresponding branching ratio,
we suggest a new analysis based
on the decay of the $\eta_b \to D^* D^{(*)}$ meson pairs. Thanks to
secondary vertex trigger capabilities of CDF and to the high resolution
achievable in the invariant mass reconstruction ($\sim 20$ MeV),
this could turn out to be the most promising search channel 
at the Tevatron.


According to the NRQCD factorization approach~\cite{Bodwin:1995jh}, 
the inclusive cross section for the $\eta_b$ in $p\bar p$ 
collisions can be written as:
\begin{eqnarray}
&&\sigma( p \bar p \to \eta_b + X) =  \nonumber\\
&&\sum_{i,j} \int dx_1 dx_2 f_{i/p} f_{j/\bar p}\,
\hat \sigma(ij \to \eta_b)\,, 
\end{eqnarray}
where
\begin{eqnarray}
&&\hat \sigma(ij \to \eta_b) = \sum_n
C^{ij}_n \langle 0| {\cal O}^{\eta_b}_n|0 \rangle\,.
\end{eqnarray}
The short-distance coefficients $C^{ij}_n$, calculable in 
perturbative QCD,  describe the
production of a quark--antiquark pair state with quantum number $n$,
while the $\langle 0 | {\cal
O}^{\eta_b}_n |0 \rangle$ are the non-perturbative matrix elements that
describe the subsequent hadronization of the $b\bar b$ pair into the
physical $\eta_b$ state.  These matrix elements 
can be expanded in powers of $v^2 \simeq
0.1$, the relative velocity of heavy quarks in the bound state, so
that, to a given accuracy, only a few terms need to be included 
in the sum over $n$.

The case of $\eta_b$ production is particularly
simple.  Compared to the leading contribution, the
$^1S_0^{[1]}(b\bar b)$, the color octect terms,
($^1S_0^{[8]},^3S_1^{[8]},^1P_1^{[8]}$), are all suppressed by $v^4$,
while the corresponding short distance coefficients start at least at
$\alpha_S^2$ as the singlet contribution~\footnote{This is at variance
with the $J/\psi$ or $\Upsilon$ production where the $C^{gg}$
coefficient is ${\cal O}(\alpha_S^3)$ and therefore color-octet
contributions can be important.}. Hence, given that the non-perturbative 
matrix elements for the singlet production are extracted from 
$\Upsilon$ decays, 
there are no unknown parameters (up to corrections of ${\cal O}(v^4)$) 
entering the estimate for $\eta_b$ production.

\begin{figure}[t]
\begin{center}
\epsfxsize=8cm 
\epsfbox{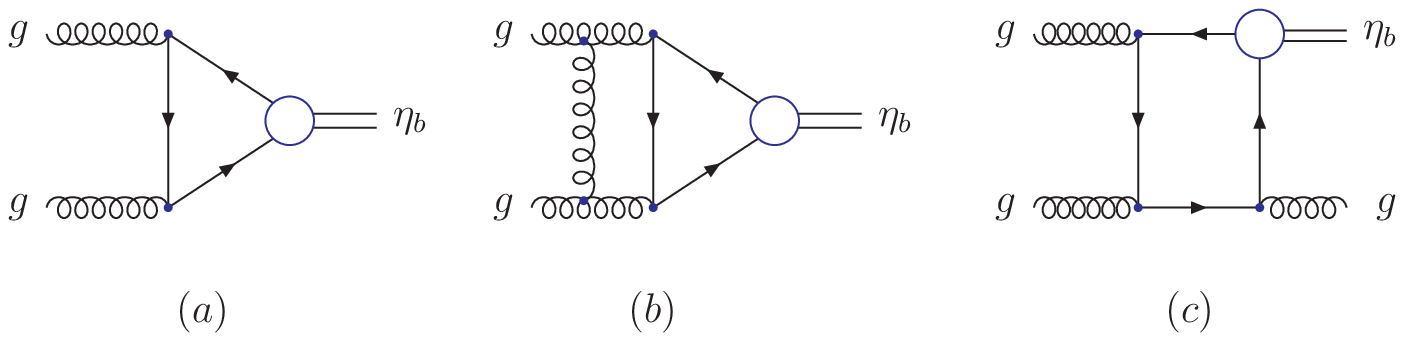}
 \vspace*{-.9cm}
\end{center}
\caption{Representative Feynman diagrams for $\eta_b$ hadro-production
at LO (a), and  virtual (b) and real (c) contributions at NLO.}

\label{fig:production}
\end{figure}

The leading-order cross section 
for $gg \to \eta_b$,  Fig.~\ref{fig:production}(a), is
\begin{eqnarray}
\hat \sigma(gg\to \eta_b) = 
\frac{\pi^3 \alpha_S^2}{36 m_b^3 \hat s } \delta
(1 - \frac{4 m_b^2}{\hat s}) \langle 0 | {\cal O}^{\eta_b}_1 (^1S_0) |
0\rangle.
\end{eqnarray}
Next-to-leading order corrections in the strong coupling have been
calculated in Refs.~\cite{Kuhn:1993qw, Petrelli:1998ge}.  These
include virtual corrections to the $2 \to 1$ process, 
Fig.~\ref{fig:production}(b),  and real $2\to 2$ processes, 
such as $gg\to \eta_b g,gq\to \eta_b q $, Fig.~\ref{fig:production}(c). 
The result for the total cross section in $p\bar p$
collisions at 1.96 TeV of center-of-mass energy is
\begin{eqnarray}
\sigma(p\bar p \to \eta_b + X) = 2.5 \pm 0.3 \, {\rm \mu b} \,,
\label{sigma_tot}
\end{eqnarray}
where we have adopted CTEQ5M1 parton densities, the corresponding
two-loop evolution for $\alpha_S(\mu_R)$ and $m_b=4.75$~GeV. The quoted
uncertainty has been estimated by summing (in quadrature) the
errors coming from various sources. The first is associated to the
choice of the renormalization and factorization scales. 
Varying them independently in the range $m_b <\mu_R,\mu_F < 4 m_b$ 
gives an uncertainty of $10\%$. 
For the non-perturbative matrix element we have used the determination from
$\Upsilon$ leptonic decay, $\langle \Upsilon |{\cal O}_1(^3S_1)
|\Upsilon\rangle = 3.5 \pm 0.3 \, {\rm GeV}^3 $, which can be related
to the $\eta_b$ production matrix element by using spin-symmetry and
vacuum saturation approximation, $ \langle 0 | {\cal O}^{\eta_b}_1
(^1S_0) | 0\rangle = \langle \Upsilon |{\cal O}_1(^3S_1)
|\Upsilon\rangle$, up to  ${\cal O}(v^4)$ corrections.
Other determinations coming from potential
models~\cite{Eichten:1995ch} and lattice
calculations~\cite{Bodwin:2001mk} fall in the range of the quoted
error. Finally, we also included the effect of an uncertainty on the
bottom (pole) mass of $\pm 50$ MeV on the cross section.  The strong
correlation to the non-perturbative matrix element extraction has been
exploited to reduce the uncertainty from this source. The
result, quoted in Eq.~(\ref{sigma_tot}), accounts for the ``direct''
contribution and does not include feed-downs from higher-mass states,
such as the $h_b$. 

\begin{figure}[t]
\begin{center}
\epsfxsize=7.5cm 
\epsfbox{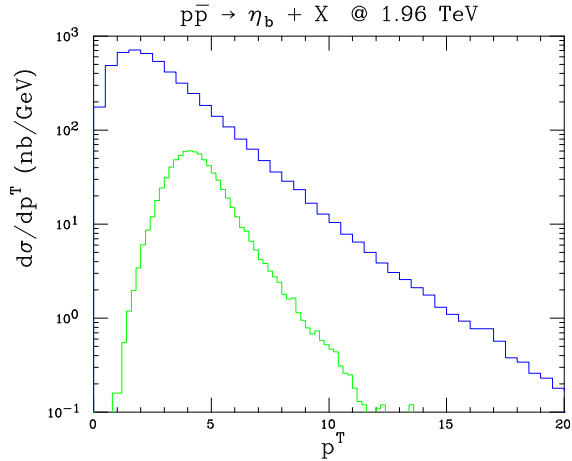}
 \vspace*{0cm}
\end{center}
\caption{Differential cross sections for $\eta_b$ production at the 
Tevatron ($p\bar p$ at 1.96 TeV). The upper curve is normalized to
the NLO total rate and describes the $p^T$ distribution for the $\eta_b$. 
The lower curve is the corresponding distribution 
for the $D$ mesons coming from the $\eta_b$ decays, 
after requiring that they both have $|\eta(D)|<1$ 
(no branching ratio included).
After the acceptance cut, the rate drops to 
15\% of the total cross section.}

\label{fig:one}
\end{figure}

For the experimental analysis it is important to know the
distribution of $\eta_b$ at small $p^T$ values. This cannot be described
accurately by a fixed-order calculation. In this region of the phase
space it is necessary to resum higher-order corrections involving
soft-gluon radiation. To this aim, we use
PYTHIA~\cite{Sjostrand:2000wi}, matched with the exact matrix elements
for $ij \to \eta_b\, k$  describing the high-$p^T$ tail, as outlined
in Ref.~\cite{Miu:1998ju}.  This procedure, which has been shown to
work well for the analogous process $gg \to H$, has the additional virtue
that it can be directly used for simulation in experiments. The results
are shown in Fig.~\ref{fig:one}, where the differential distribution in
$p^T$ for the $\eta_b$ is shown (upper curve). Hadronization and initial $k_T$ effects are not included.  The normalization of the inclusive $\eta_b$ 
distributions is obtained from the NLO result, Eq.~(\ref{sigma_tot}).

As already stated above, the cross section for the
$\eta_b$ in the central region is about four times larger than was
estimated in Ref.~\cite{Braaten:2000cm} by rescaling the
$\Upsilon$ cross section at high transverse momentum.  
This procedure is expected to provide an underestimate,  
since it assumes that the $p^T$ spectra for $\Upsilon$ and $\eta_b$ 
have a similar shape at small $p^T$. In fact $\Upsilon$ color-singlet 
production proceeds at LO through $gg \to \Upsilon g$ and 
therefore vanishes at $p^T \sim 0$,  
where the largest part of the $\eta_b$'s are produced. 

\begin{figure}[t]
\begin{center}
\epsfxsize=6cm 
\epsfbox{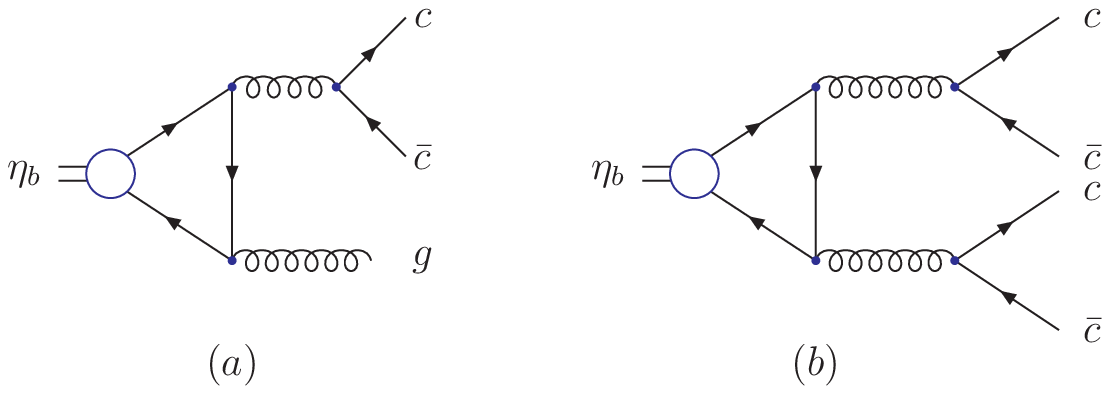}
\vspace*{-1cm}
\end{center}
\caption{Representative Feynman diagrams for the $\eta_b$ inclusive decay into
 (a) two-charm (two diagrams) and (b) four-charm (four diagrams) states.}
\label{fig:decay}
\end{figure}

With such a large number of events expected, it is interesting to
consider in detail the rare exclusive decays that might give triggerable
signatures. It is easy to show that direct decays into photons or lepton
pairs give either too small branching ratios or very difficult
experimental signatures (such as $\eta_b \to \gamma \gamma$ 
whose branching ratio is ${\cal O}(10^{-5})$).  
In Ref.~\cite{Braaten:2000cm}, the branching ratio 
${\rm Br}(\eta_b\to J/\psi J/\psi)$ was estimated 
through scaling from the
analogous $\eta_c \to \phi \phi$, finding a value compatible with
$7 \times 10^{-4\pm1}$. This is probably an
overestimate. As a simple upper bound, let us consider the inclusive
decay rate of the $\eta_b$ into four-charm states
\begin{eqnarray}
\Gamma(\eta_b \to J/\psi\,J/\psi) < \Gamma(\eta_b \to c \bar c c \bar c).
\end{eqnarray}
The inclusive rate can be calculated at leading order 
by considering the four Feynman diagrams, 
such as the one shown in Fig.~\ref{fig:decay}(b). The result is 
\begin{eqnarray}
&& {\rm Br}(\eta_b \to c \bar c c \bar c) = 
1.8^{+2.3}_{-0.8} \times 10^{-5}  \,. 
\label{eq:br4c}
\end{eqnarray}
where  $m_c=1.45$ GeV, $m_b=4.75$ GeV, $\alpha_S(2 m_b)=0.182$
and the NLO expression for the inclusive total
width~\cite{Hagiwara:1981nv,Petrelli:1998ge} have been used.  
The amplitudes have been calculated analytically while the 
integration over the phase space has been performed numerically.  
The uncertainty has been estimated by varying the renormalization 
scale between $m_b < \mu_{\rm R}<4 m_b$ and the masses in the 
$\pm 50$ MeV range. The four-charm branching ratio is very sensitive 
to the value of the charm mass, which dominates its uncertainty.
The above result shows that the inclusive rate is already smaller
than the lower bound obtained in Ref.~\cite{Braaten:2000cm}; further
suppression is expected mainly because many other decay
modes to charmed mesons (or other charmonium states) should contribute
to the saturation of the inclusive rate. 

Given the result in Eq.~(\ref{eq:br4c}), 
a comment on the reliability of the estimate based on the scaling from
${\rm Br}(\eta_c\to \phi \phi)$ is in order.
To this aim we recall that the  decay of a scalar $Q\bar{Q}$ meson 
into two vector states is suppressed in perturbative QCD~\cite{Brodsky:1981kj}.
A non-trivial check of this selection rule is that 
the branching ratio ${\rm Br}(\eta_b\to J/\psi J/\psi)$ 
is exactly zero when calculated at LO in 
the NRQCD double expansion in $\alpha_S$ and $v^2$.
For the same reason one would expect the rate of $\eta_c\to\phi\phi$
to be suppressed, in contradiction with the measured value of about 1\%.
This entails that some other (non-perturbative) mechanism is responsible
for this decay process~\cite{Anselmino:1990vs}. 
Rescaling by $(m_c/m_b)^4$ the branching ratio of
$\eta_c\to\phi\phi$ to obtain the branching ratio of $\eta_b\to J/\psi
J/\psi$ amounts to rescaling by the same factor also the effect 
of non-perturbative or higher-order contributions 
that are likely to be crucial in determining the $\eta_c$ decay, 
but less and less important as we pass from the $\eta_c$ 
to the $\eta_b$ system.

From the phenomenological point of view, Eq.~(\ref{eq:br4c})
implies ${\rm Br}(\eta_b\to \mu^+ \mu^- \mu^+ \mu^-)  < 6 \times 10^{-8}$,
which makes a search in this channel very hard.
As an alternative, we propose to consider the decays into charmed mesons,
$\eta_b \to D^{*} D^{(*)}$ \footnote{The decay
$\eta_b \to D D$ is forbidden by parity conservation.}. 

A computation of the exclusive decay rates is missing and
does not seem to be feasible within the framework of
quark models or QCD sum rules. Probably, there is room to face such
a problem using lattice techniques.
Here we assume that the exclusive decays into $D^* D^{(*)}$ dominate
the inclusive rate into charm:
\begin{eqnarray}
\Gamma(\eta_b \to D^* D^{(*)} ) \lesssim \Gamma(\eta_b \to c \bar c +X ).
\end{eqnarray} 
We find that the largest contribution to the decay of the $\eta_b$ into
charmed states is given, Fig.~\ref{fig:decay}(a), by
\begin{eqnarray}
&&  {\rm Br}(\eta_b \to c \bar c g) =1.5^{+0.8}_{-0.4}\%  \,,
\label{eq:br2c}
\end{eqnarray}
where we used the same input parameters and estimated the uncertainties 
as in the decay into the four-charm states. 
It can now be argued that not much suppression is expected in passing
from the inclusive to the exclusive decays. For example, decay rates
into charmonium states should be of the same size 
$\Upsilon \to J/\psi +X $, 
{\it i.e.}, ${\cal O}(10^{-3}$)~\cite{Hagiwara:2002fs}. 
Emission of extra pions should also be considered, 
but some of these contributions would be 
automatically included in the experimental analysis, 
{\it e.g.}, $\eta_b \to D D^* \to D  D \pi$, as 
non-resonant diagrams. However, to be conservative, 
we consider the range 
$10^{-3}<{\rm Br}(\eta_b \to D^* D^{(*)} )<10^{-2}$
in the following phenomenological analysis.

There are other decay modes leading to a two-charm
final state that have not been included in our estimate of the
branching ratio.  One is the decay of $\eta_b \to g^* g^* \to c \bar
c$ through a (box) loop, which proceeds at order $\alpha_S^4$ and is
further suppressed by loop factors.  Another is the decay $\eta_b \to
g^* \to c \bar c$, via a $^3S_1^{[8]}$ state, which is only,
$\alpha_S^2$ but it is suppressed by the color octet matrix element
$\langle \eta_b |{\cal O}_8(^3S_1) |\eta_b \rangle$.  Assuming a
scaling ${\cal O}(v^4)$ for the non-perturbative matrix element, this
process gives a non-negligible contribution to the branching ratio,
about $5\times 10^{-3}$. However, this result is affected by a large
uncertainty and could be much smaller as suggested by
studies~\cite{Maltoni:1998nh,Fleming:2002rv}
about the size of color
octect matrix elements in $\Upsilon$ decays.


Results for the $p^T$ distribution of the $D$ mesons 
from the $\eta_b$ decay are shown in Fig.~\ref{fig:one} (lower
curve). No branching ratio is included, the difference in rate coming only
from the requirement that both $D$ mesons be central, $|\eta(D)|<1$. 
As expected, the $p^T$ distribution peaks just below $M_{\eta_b}/2$.
The efficiency for the geometrical acceptance of the detector is 
found to be about 15\%. By adding the requirement that at least one 
$D$ meson has $p^T>5$ GeV, one is left with only 
4\% of the total number of events produced. 
The above efficiency can be folded with our 
estimate of the branching ratio leading to 
$10^4-10^5$ $D^* D^{(*)}$ triggerable events expected 
from the $\eta_b$ decay  in 100 pb$^{-1}$ of integrated luminosity 
at the Tevatron. The final number of reconstructed events will depend 
on the decay modes of the $D$ and $D^*$ 
mesons and on the associated experimental efficiencies. 
We leave this to more detailed experimental studies and 
only add a few comments. First we note that, 
according to the arguments outlined above, 
perturbative QCD predicts the $\eta_b\to D^*D^*$ decay to
have a smaller rate with respect to $\eta_b\to D^* D$.
In this case, it is reasonable to expect that different 
charge assignments, such as $D^{*0} D^0, D^{*+} D^-, D^{*-} D^+$, 
will occur with the same probability of $\frac{1}{3}$. 
Finally, we recall that the cleanest 
signatures have small branching ratios, ${\rm Br}(D^0 \to K^- \pi^+)=3.90\%$ 
and ${\rm Br}(D^+ \to {\overline K^0} \pi^+)=2.77 \%$~\cite{Hagiwara:2002fs},
leading to a factor of about $10^{-3}$ drop in the rate if both $D$ mesons
are required to decay through these channels. We foresee that sizeable 
improvements could be achieved by requiring that just one of
two $D$ mesons decays through a very clean signature, providing  
an efficient trigger. 


To summarize, we have presented the NLO QCD, prediction for $\eta_b$
production at the Tevatron, including a resummed result for the $p^T$
spectrum obtained with a dedicated implementation in PYTHIA. The
production rate is large, of the order of a few $\mu$b, and allows 
for the search of the $\eta_b$ through rare exclusive decays.  
We argued that the branching ratio into $J/\psi\, J/\psi$ is 
probably too small and we suggested to look for
the $\eta_b$ through its decay into $D^* D^{(*)}$ mesons. 
Our results indicate that the $\eta_b$ 
could be eventually observed during Run II at  the Tevatron.

\begin{acknowledgments}

We have greatly benefited from conversations with M.~Campanelli.  We
are grateful to T.~Sjostrand for his help with PYTHIA. We thank
M.~Greco and S.~Willenbrock for their comments on the manuscript. 
ADP thanks T.~Feldmann and G.~Nardulli for informative discussions.
FM warmly thanks the Department of Physics of the 
Terza Universit\`a di Roma for the kind hospitality and support.

\end{acknowledgments}

\bibliography{database}
\end{document}